\documentclass[twocolumn,preprintnumbers,amsmath,amssymb]{revtex4}  

\usepackage{graphicx}
\usepackage{dcolumn}
\usepackage{bm}
\usepackage{multirow}
\usepackage{color}
\usepackage{amsmath}

\begin{document}
\title{Exciton states in monolayer MoSe$_2$: impact on interband transitions}

\author{G. Wang}
\author{I. C. Gerber}
\author{L. Bouet}
\author{D. Lagarde}
\author{A. Balocchi}
\author{M. Vidal}
\author{E. Palleau}
\author{T. Amand}
\author{X. Marie}
\author{B. Urbaszek}
\affiliation{%
Universit\'e de Toulouse, INSA-CNRS-UPS, LPCNO, 135 Av. de Rangueil, 31077 Toulouse, France}


\begin{abstract}
We combine linear and non-linear optical spectroscopy at 4~K with ab initio calculations to study the electronic bandstructure of MoSe$_2$ monolayers. In 1-photon photoluminescence excitation (PLE) and reflectivity we measure a separation between the A- and B-exciton emission of 220 meV. In 2-photon PLE we detect for the A- and B-exciton the 2p state 180~meV above the respective 1s state. In second harmonic generation (SHG) spectroscopy we record an enhancement by more than 2 orders of magnitude of the SHG signal at resonances of  the  charged exciton and the 1s and 2p neutral A- and B-exciton. Our post-Density Functional Theory calculations show in the conduction band along the $K-\Gamma$ direction a local minimum that is energetically and in k-space close to the global minimum at the K-point. This has a potentially strong impact on the polarization and energy of the excitonic states that govern the interband transitions and marks an important difference to MoS$_2$ and WSe$_2$ monolayers. 
\end{abstract}


                             
\maketitle

\section{Introduction}
\label{sec:intro}
\begin{figure}[h]
\includegraphics[width=0.47\textwidth]{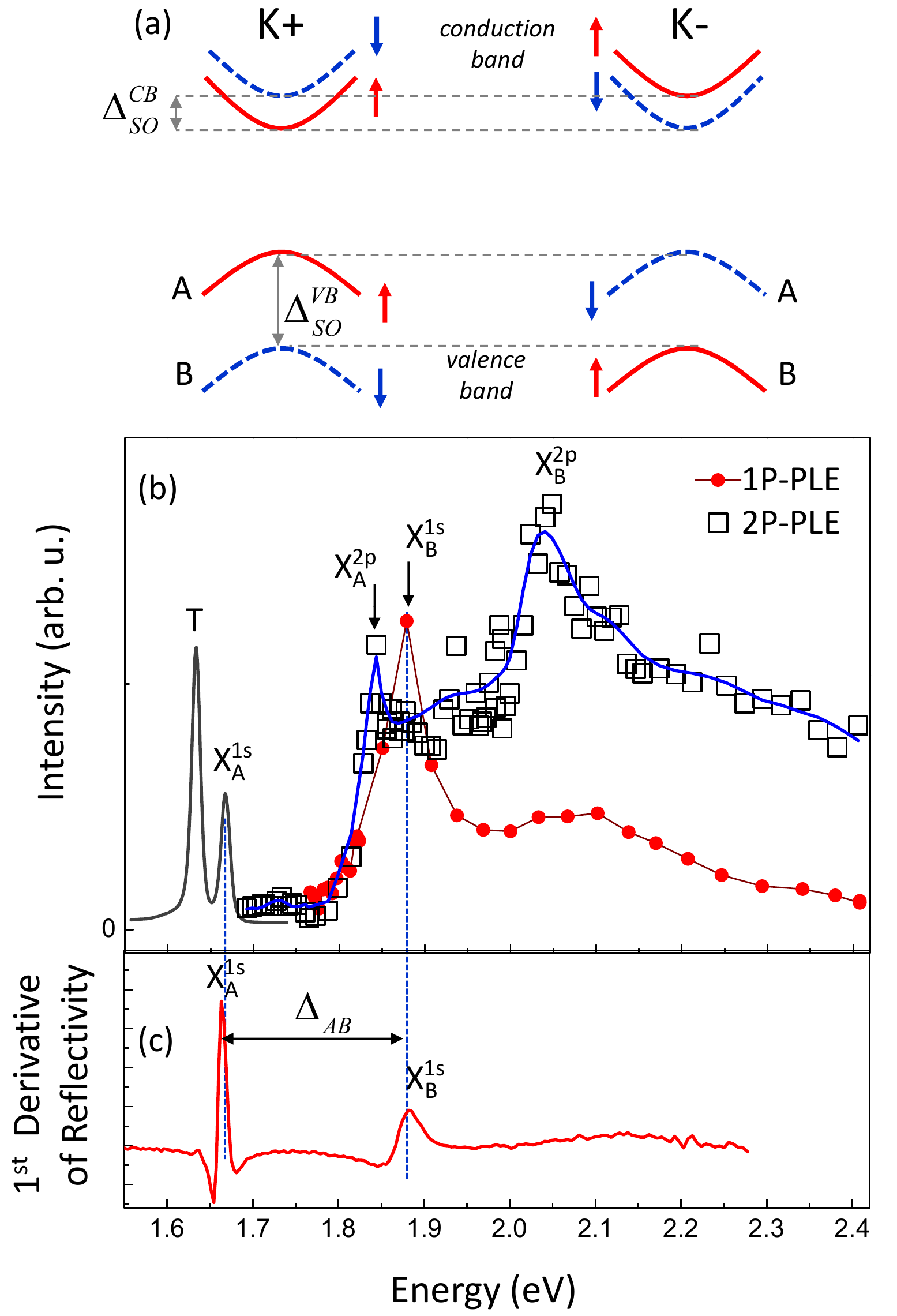}
\caption{\label{fig:fig1} (a) Conduction (CB) and valence states (VB) in a single particle picture. (b) The intensity of the neutral 1s A-exciton PL (shown as gray solid line) is recorded as a function of laser energy. In 1-photon PLE (red circles) the 1s B-exciton state (X$_B^{1s}$) is identified, in 2-photon PLE (open black squares) the peaks are assigned to the  2p A-exciton (X$_A^{2p}$) and 2p B-exciton (X$_B^{2p}$). (c) Maxima of the first derivative of the reflectivity allow to assign the 1s A-exciton and B-exciton state energies.
}
\end{figure}
Monolayers (MLs) of the transition metal dichalcogenides (TMDCs) MoS$_2$, MoSe$_2$, WS$_2$ and WSe$_2$ (abbreviated MX$_2$) are semiconductors with a direct bandgap in the visible region \cite{Mak:2010a,Splendiani:2010a,Zhang:2014a}. MX$_2$ MLs are promising materials for (opto-)electronics \cite{Geim:2013a,Butler:2013a,Zhao:2013b,Mak:2010a,Splendiani:2010a,Radisavljevic:2011a,Ross:2014a}, non-linear optics \cite{Kumar:2013a,Zeng:2013a,Li:2013b,Yin:2014a,Malard:2013a,Trolle:2014a,Wang:2015b,Gruning:2014a}  and for exploring electron k-valley physics \cite{Xiao:2010a,Mak:2014a,Xu:2014a}. The MX$_2$ ML materials share common characteristics: (i) Their optical properties are dominated by excitons, strongly Coulomb-bound electron hole pairs \cite{Cheiwchanchamnangij:2012a,Komsa:2012a,Ross:2013a,Song:2013a,He:2014a,Ugeda:2014a,Chernikov:2014a,Ye:2014a,Wang:2015b,Klots:2014a,Zhu:2014a,Hanbicki:2015a,Hill:2015a}. (ii)  In these TMDC MLs crystal inversion symmetry breaking together with the strong spin-orbit (SO) interaction leads to a coupling of carrier spin and k-space valley physics, i.e., the circular polarization ($\sigma^+$ or $\sigma^-$) of the absorbed or emitted photon can be directly associated with selective carrier excitation in one of the two non-equivalent K valleys (K$^+$ or K$^-$, respectively) \cite{Xiao:2012a, Cao:2012a,Mak:2012a,Sallen:2012a,Kioseoglou:2012a,Jones:2013a,Mak:2014a}. \\
\indent Using non-resonant laser excitation in ML MoS$_2$  \cite{Mak:2012a,Zeng:2012a,Kioseoglou:2012a,Sallen:2012a}, WSe$_2$ \cite{Jones:2013a,Wang:2015b} and WS$_2$ \cite{Zhu:2014b} high values  ($\approx 50\%$) for the circular polarization $P_c$ of the stationary photoluminescence (PL) corresponding to successful valley polarization have been reported. In contrast, the polarization reported for the promising material MoSe$_2$ \cite{Singh:2014a,Kumar:2014a,Ross:2013a} under similar experimental conditions is surprisingly very low ($\lesssim 5\%$) \cite{Macneill:2015a,Li:2014a,Wang:2015a}. Experiments combing PL excitation (PLE) and time resolved PL suggest either ultra fast polarization relaxation in MoSe$_2$ in the sub-picosecond range or inefficient optical polarization generation due to anomalies in the bandstructure to be at the origin of this low valley polarization \cite{Wang:2015a}. \\
\indent Our target is to investigate how an eventual bandstructure anomaly (local extremum),  can influence the nature of the optical transitions. To this aim we combine linear and non-linear optical spectroscopy at T=4~K with calculations beyond standard Density Functional Theory (DFT) to study the electronic bandstructure of MoSe$_2$ MLs. In 2-photon PLE we detect 180 meV above the 1s state the 2p state of the A-exciton, well separated in energy from the 1s B-exciton emission recorded in both reflectivity and 1-photon PLE.  The energy position of the neutral and charged exciton transitions is determined using second harmonic generation (SHG) spectroscopy, as the SHG signal is enhanced by more than 2 orders at these particular resonances, important for applications in non-linear optics.\\
\indent The strong excitonic effects that dominate the 1- and 2-photon spectroscopy results in the investigated MoSe$_2$ MLs have important consequences not just in terms of the transition energy \cite{Ugeda:2014a}.  Due to the small exciton Bohr radius, the extension in k-space is considerable and the exciton wavefunction will include contributions from states far away from the K-point of the Brillouin zone where the direct free carrier (electronic) bandgap of the MX$_2$ MLs is situated. The position in energy and in k-space of the adjacent local minima in the conduction band (CB) and maxima in the valence band (VB), respectively can be very different from one MX$_2$ material to another \cite{Kormanyos:2013a,Kormanyos:2014b}.  Our DFT-GW calculations show in the CB of ML MoSe$_2$ a local minimum that is energetically and in k-space close to the global minimum at the K-point \cite{Bradley:2015a}. We evaluate how the proximity of this local minimum along the $K-\Gamma$ direction can influence the polarization and energy of the excitonic states that govern the interband transitions, marking an important difference compared to MoS$_2$ and WSe$_2$ monolayers.
\begin{figure} [h]
\includegraphics[width=0.49\textwidth]{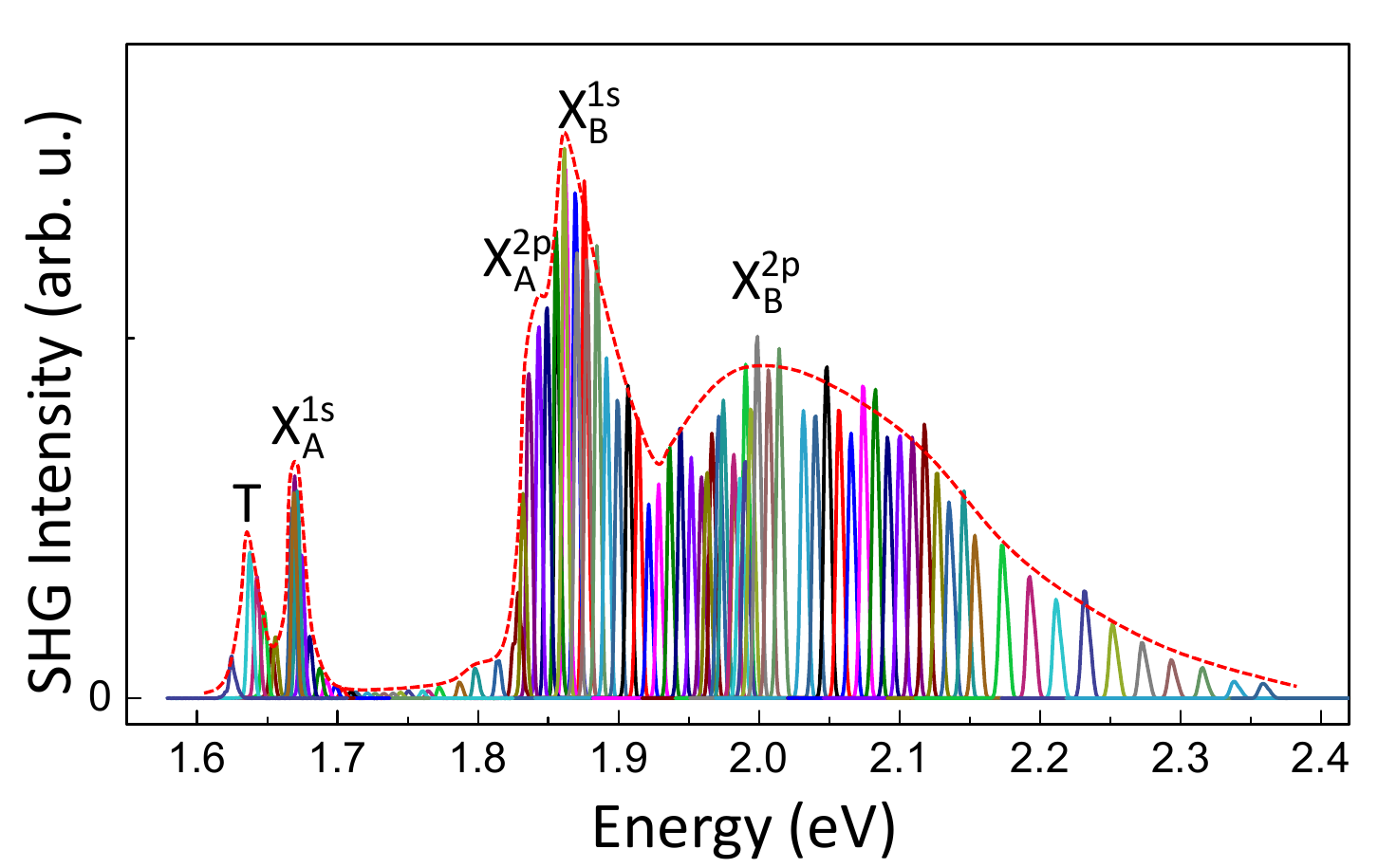}
\caption{\label{fig:fig2} The second harmonic generation (SHG) signal is plotted as a function of twice the laser photon energy. The local maxima corresponding to the neutral and charged exciton states are marked (see text). 
}
\end{figure} 
\section{Spectroscopy of Exciton States}
\label{sec:spectro}
\indent Similar to the hydrogen model, an electron $–$ hole pair in the TMDC ML interacts through an attractive Coulomb potential 
and will form below the gap a series of excitonic Rydberg-like states with
definite parity, where $ns$ states are of even, and $np$ states are of odd parity (n is an integer) \cite{Klingshirn:2006a}.
The exciton binding energy $E_b$ of the order of 0.5~eV in ML MoSe$_2$ can be estimated by determining the free carrier bandgap in scanning tunnelling
spectroscopy (STS) and subtracting the neutral exciton energy (corresponds to the 1s A-exciton state), see \textcite{Ugeda:2014a}.
Here we aim to uncover in addition to the 1s exciton ground state also higher excited exciton states, which were shown to dominate the linear and non-linear optical response in MoS$_2$ and WSe$_2$ MLs \cite{Wang:2015b,Carvalho:2015a}. In Fig.~\ref{fig:fig1}b we see a typical time-integrated PL spectrum at T=4~K for ML MoSe$_2$ with two prominent peaks (see Appendix for information on samples and the experimental set-up). The higher energy peak (FWHM=10~meV) at 1.667~eV has previously been attributed to the neutral A-exciton X$_A^{1s}$ \cite{Ross:2013a,Li:2014a,Macneill:2015a}. At 1.633~eV we record the trion emission (T) corresponding to a binding energy of 34~meV, in agreement with previous measurements \cite{Ross:2013a,Li:2014a,Macneill:2015a,Wang:2015a}. 
\subsection{One-Photon PLE experiments}
\label{sec:1pple}
\indent Here we detect the intensity of the X$_A^{1s}$ PL emission as a function of the laser energy, shown in Fig.~\ref{fig:fig1}b.
We find a clear maximum at 1.885~eV, which we attribute to the 1s B-exciton state. In order to confirm this, we have performed reflectivity measurements using a white light source. We find in reflectivity two prominent transitions in Fig.~\ref{fig:fig1}c, one at 1.667~eV, which corresponds exactly to the neutral A-exciton emission energy measured in PL. This indicates negligible localization of excitons in this sample (no Stokes shift). The second transition in reflectivity at 1.885~eV gives the position of the 1s B-exciton. The measured difference between 1s A and B-exciton is $\Delta_{AB}\simeq 220$~meV determined from both PL and reflectivity. From our DFT calculations (see section \ref{sec:calcul} for details) we obtain a spin splitting of $\Delta_{SO}^{VB}=183$~meV in the valence band. This is in very good agreement with the measured value of 180~meV from angle-resolved photoemission spectroscopy (ARPES) measurements at T=40~K \cite{Zhang:2014a}. The CB spin splitting is predicted to have the opposite sign compared to the VB, as sketched in Fig.~\ref{fig:fig1}a \cite{Liu:2013a,Kosmider:2013a,Kormanyos:2014b}. We obtain from our DFT calculations a CB spin splitting of $\Delta_{SO}^{CB}=-30$~meV. The measured difference $\Delta_{AB}$ of 220~meV is in close agreement with the calculated energy difference taking into account the VB and CB spin splitting of 183+30=213~meV. In addition, we find in DFT calculations in section \ref{sec:calcul} that the effective masses in CB and VB are slightly different for different spin states, and hence also the exciton binding energies of A- and B-excitons are expected to differ and contribute to the measured value of $\Delta_{AB}$.
\subsection{Two-Photon PLE experiments}
\label{sec:2pple}
\indent An experimental signature of excited exciton states, in analogy with the $ns$ and $np$ states of the hydrogen atom, have not been reported yet for MoSe$_2$ MLs, to the best of our knowledge. The excited exciton states and the exact exciton binding energy $E_b$ in MX$_2$ MLs are currently debated in the literature for WS$_2$ \cite{Chernikov:2014a,Ye:2014a,Zhu:2014a,Hanbicki:2015a}, MoS$_2$ \cite{Klots:2014a,Hill:2015a} and WSe$_2$ \cite{He:2014a,Wang:2015b,Hanbicki:2015a}, with differences in the reported $E_b$ for the same material by factor of 2, so additional experiments are important. In our 2-photon PLE measurements we can directly address exciton states with p-symmetry and not with s-symmetry as in the 1-photon PLE and reflectivity. In the 2-photon PLE of Fig.~\ref{fig:fig1}b we find a well defined peak at 1.844~eV, which we assign to the 2p A-exciton transition, not obscured by the 1s B-exciton, which is parity forbidden here. We measure an energy difference 1s-2p of 180~meV. Using a binding energy of the order of $E_b\approx 0.5$~eV \cite{Ugeda:2014a} in a simple 2D hydrogen model \cite{Klingshirn:2006a}, the energy separation 1s-2s,2p is expected to be $\frac{8}{9}E_b\simeq 440$~meV, much larger than our measured value of 180~meV. This shows for ML MoSe$_2$ a strong deviation from a simple hydrogenic series, observed also in ML WS$_2$ and WSe$_2$, very likely due to a strong variation of
the effective dielectric constant as a function of the spatial extension of the exciton state \cite{Cudazzo:2011a,Deslippe:2009a}.  Interestingly we observe a second, clear maximum at 2.06~eV, about 175~meV above the B-exciton 1s state. We tentatively assign this peak in 2-photon PLE to the 2p B-exciton state. 
\subsection{Second Harmonic Generation Spectroscopy}
\label{sec:SHG}
\indent In addition to the 2-photon PLE signal we can also plot the intensity of the SHG signal as a function of Laser energy in Fig.~\ref{fig:fig2}. The SHG spectroscopy results in this ML material without crystal inversion centre also show clear resonances at the exciton energies. Interestingly we observe a strong SHG signal at the transition energy of the charged exciton (trion). This hints at a substantial density of states of this complex. \\
\indent The SHG signal at the 1s B-exciton is about 120 times higher than in the region between 1.7 and 1.8~eV. This finding shows that excitonic resonances dominate the non-linear optical response of ML MoSe$_2$. The SHG signal is non-zero for all laser energies, the SHG peaks outside the exciton resonances between 1.7 and 1.8~eV are not clearly visible on the linear scale used in Fig.~\ref{fig:fig2}.  We confirm the energy position of the 1s and 2p exciton levels A and B. Note that resonances of 1s exciton states in SHG are forbidden if strict electric dipole selection rules apply. The prominent 1s features in our SHG spectrum in Fig.~\ref{fig:fig2} can come from the interplay of electric dipole with magnetic dipole transitions \cite{Wang:2015b,Lafrentz:2013a}. In addition strict electric dipole selection rules could be slightly relaxed if the overall symmetry of the crystal is lowered by extrinsic effects such as the substrate.
\section{Electronic Band-Structure calculations}
\label{sec:calcul}
\indent In the experiments on ML MoSe$_2$ we find sharp excitonic features that are well defined as for the related ML material WSe$_2$ \cite{Wang:2015b}. Despite similar exciton binding energy and excited state spectrum in ML WSe$_2$ and ML MoSe$_2$ the optical valley polarization using non-resonant laser excitation can be generated in the former but not in the latter \cite{Li:2014a,Macneill:2015a,Wang:2015a}. One reason could be an anomaly in the bandstructure, which we aim to uncover in band structure calculations for MoSe$_2$. For comparison and to validate our computational approach we have performed in parallel calculations with the more thoroughly studied ML MoS$_2$. The bandstructure of ML MoSe$_2$ and MoS$_2$ can be compared in Fig.~\ref{fig:fig3}, where striking differences in both valence and conduction band appear. We extract the effective carrier masses, the band gaps and the exciton binding energies. We discuss the most important features for optical transitions in detail, in particular the competition between direct and indirect exciton states in MoSe$_2$ MLs. \\
\subsection{Quasi-Particle Band Structure}
\label{sec:bands}
\begin{figure*}[ht!]
\includegraphics[width=0.9\textwidth]{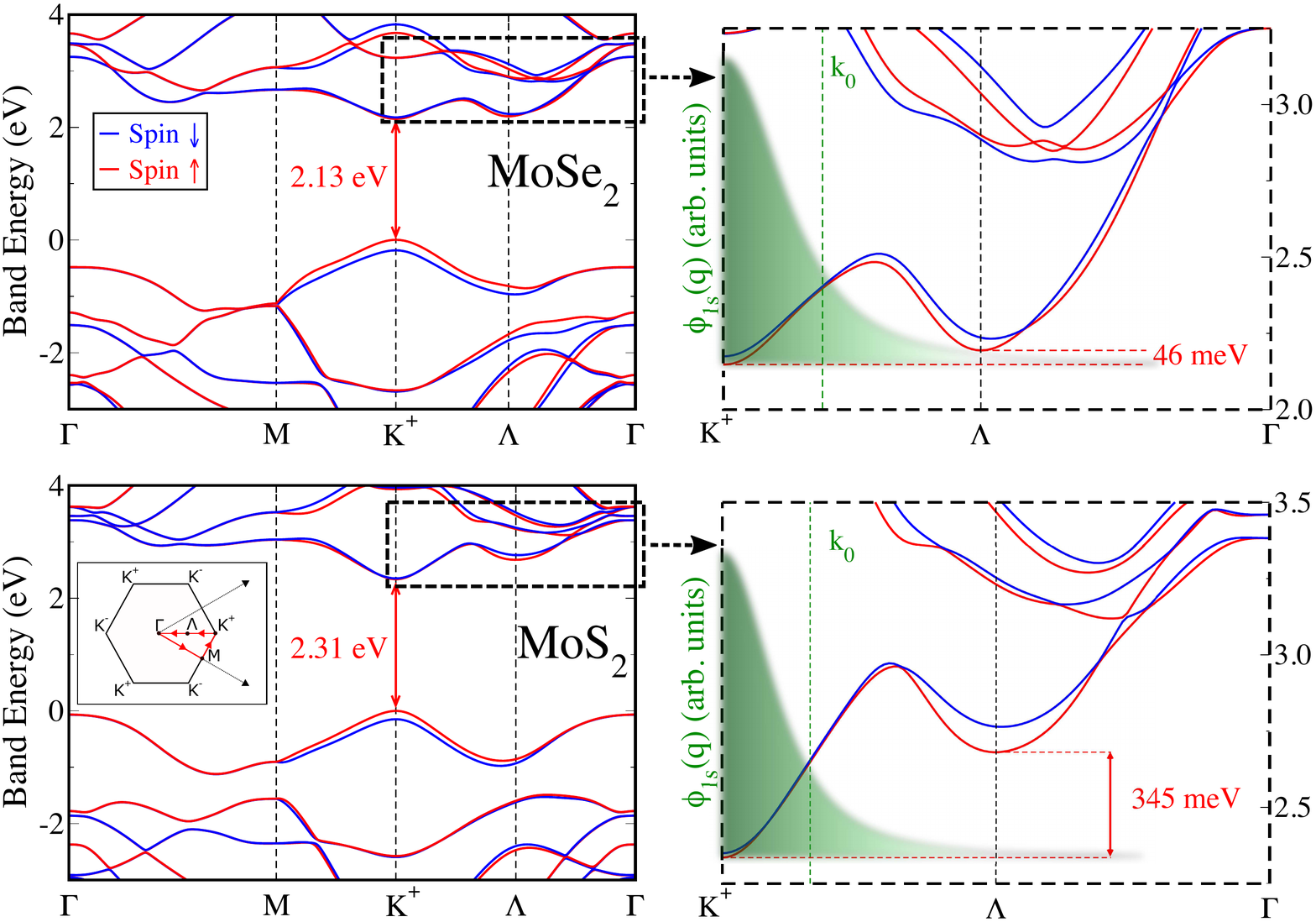}
\caption{\label{fig:fig3}  $G_0W_0$ band structure for freestanding MoSe$_2$ ML (top left). Zoom at K point and along K-$\Gamma$ direction (top right).  MoS$_2$ ML results are in the bottom panels. Zero energy represents the Valence Band Maximum. The green shaded area is a schematic representation of the extension in k-space of the 1s A-exciton wavefunction, see. Eq.~(4) in the text.
}
\end{figure*} 
\indent The comparison of quasi-particle band structure calculated by DFT (see Appendix for computational details) of ML MoS$_2$ and MoSe$_2$ provides interesting insights: when replacing S with Se, the spin-splitting is enhanced. For MoS$_2$  the valence and conduction state the spin-splittings at the K$^{+}$ point are $\Delta_{SO}^{VB}=143$~meV and $\Delta_{SO}^{CB}=-13$~meV, respectively. For ML MoSe$_2$ we find $\Delta_{SO}^{VB}=183$~meV and $\Delta_{SO}^{CB}=-30$~meV. The direct band gap values (neglecting excitonic effects) at K$^{+}$, depicted by arrows in the left panels of Fig.~\ref{fig:fig3}, are 2.13 and 2.31 eV for MoSe$_2$ and MoS$_2$, respectively. 
The latter value is lower than previous theoretical studies as energies from 2.41 to  2.97 eV have been reported~\cite{Komsa:2012a,Cheiwchanchamnangij:2012a,Ramasubramaniam:2012a,MolinaSanchez:2013,Qiu:2013a,Shi:2013a,Klots:2014a}. It is important to take into account that the uncertainty in this type of band gap calculations is in the hundreds of meV range, depending on the computational settings. The obtained results critically depend on the choice of the $GW$ methods used (self-consistent or partially self-consistent scheme), the number of unoccupied states included, the vacuum height and the k-point sampling.  
Interestingly using exactly the same computational settings, our MoSe$_2$ direct band gap estimate is closer to recent reports:  2.33 eV for a $G_0W_0$ calculation in Ref.\onlinecite{Horzum:2013a},  when Ugeda \textit{et al}~\cite{Ugeda:2014a} propose 2.26 eV for a free standing MoSe$_2$ ML. On the experimental side, using STS techniques the band gap of MoSe$_2$ ML is 2.18~$\pm$ 0.04~eV~\cite{Bradley:2015a} but it includes substrate screening effects.\\
\indent If one compares the full band-structure of the two material systems another striking difference appears: in MoS$_2$ ML, the topmost valence bands in $\Gamma$ and K$^{+}$ are only separated by 65 meV.  This  feature appears to be essential in the mechanism of direct to indirect band gap transition and its evolution with layer thickness (1ML, 2ML ...)~\cite{Jin:2013a} and is expected to have a strong impact on the optically generated valley polarization as a function of laser energy \cite{Kormanyos:2013a,Lagarde:2014a}. The same trend for the gap evolution with layer thickness applies to  MoSe$_2$ \cite{Zhang:2014a}, albeit with a much larger $\Gamma$ to K$^{+}$ energy separation of 485 meV for 1ML. \\
\indent For ML MoSe$_2$ our calculations show a remarkable anomaly in the CB, which could play a key role for the valley polarization dynamics: In ML MoSe$_2$ there is a small energy difference of only 46~meV between the CB minimum in K$^+$ and in $\Lambda$, as can be clearly seen in Fig.~\ref{fig:fig3}. The corresponding value for MoS$_2$ is 345~meV and therefore not in competition with direct optical transitions at the K points. We note another important difference: The $\Lambda$ position along the K$-\Gamma$ line for MoS$_2$ lies exactly at the $(1/6,1/6,0)$ coordinates. In contrast, for MoSe$_2$ it is slightly shifted towards the K point. Interestingly the spin-up and spin-down minima are not positioned at the same k-value. Note that the appearance of this local minimum has also been predicted recently in \cite{Bradley:2015a}.\\
\indent In the following we discuss how the proximity of the CB minimum at $\Lambda$ can potentially influence the optical transitions. Excitons with large binding energies in ML MoSe$_2$ have a very small Bohr radius and as a result extend in k-space well beyond the K-point (see Fig.~\ref{fig:fig3}). To allow for a more quantitative discussion, we need to extract the effective carrier masses from our calculated band structure. At VB and CB extrema, effective masses for holes and electrons can be extracted with a simple quadratic fitting procedure. The extracted mass values will depend on the selected k-value range around the extrema. For consistency,  we have used the interval $[-k_0/2,+k_0/2]$, with $k_0$ being defined as $1/(2\times a_B^{2D})$, where $a_B^{2D}$ is the Bohr radius of the exciton ground state in a Wannier-Mott picture of electron-hole pair \cite{Ye:2014a}. The computed effective mass values are given in Table \ref{tab1}. For MoS$_2$, the $m_c$ and $m_v$ values are only slightly different from previous studies~\cite{Cheiwchanchamnangij:2012a,Ramasubramaniam:2012a, Shi:2013a}, the difference being reduced in the electron-hole pair effective mass $\displaystyle \mu=\frac{m_c~m_v}{m_c+m_v}$ values. There is a certain spread in values, especially for the MoS$_2$ hole effective mass with the value provided in Ref.~\onlinecite{Qiu:2013a}, or for MoSe$_2$ values~\cite{Horzum:2013a}. We obtain smaller values, but we attribute these differences to the choice of the extension in k-space around the extrema. Indeed, if one takes a fitting interval of length $2 k_0$ the effective masses are significantly increased by 30\% in both material systems. Interestingly spin-up and spin-down excitons have different effective masses at K$^+$: 0.21~$m_0$ \textit{vs} 0.24~$m_0$ for $\mu_{\uparrow}$ and $\mu_{\downarrow}$ respectively  for MoS$_2$, 0.25~$m_0$ and 0.28~$m_0$ in MoSe$_2$, due to larger values of the spin-down $m_c$ and $m_v$, in both systems. This confirms that the exciton binding energies for A- and B-excitons should be slightly different, which will in turn influence the energy difference $\Delta_{AB}$ between the corresponding optical transitions measured in Fig.~\ref{fig:fig1}. \\
\indent On top of $G_0W_0$ calculations, exciton binding energies $E_b$, given in Table \ref{tab1}, have been extracted from the imaginary part of transverse dielectric constant after solving the Bethe-Salpeter Equation (BSE). Our 0.58 eV estimate is in reasonable agreement with previous theoretical studies for the MoS$_2$ A-exciton binding energy: values from 0.55~\cite{Huser:2013a}, around 0.6~\cite{Shi:2013a, Klots:2014a}, or around 0.9~\cite{Cheiwchanchamnangij:2012a,Qiu:2013a} and up to 1.1 eV\cite{Ramasubramaniam:2012a, Komsa:2012a} have been reported. Please note the computational details are extremely different for each of these values. It is therefore difficult to extract precise trends. Alternatively, the spread in values should rather be used as a reasonable error bar for this type of calculations.  For MoSe$_2$ theoretical values for $E_b$ of 0.9~\cite{Ramasubramaniam:2012a},  0.78~\cite{Komsa:2012a} and 0.65 eV~\cite{Ugeda:2014a} can be found in the recent literature. This indicates that the calculated exciton binding energy in ML MoSe$_2$ is smaller than in MoS$_2$, as observed experimentally. \\
\begin{table}[h]
\caption {Free carrier band-gap $E_g$, exciton binding energy $E_b$, CB effective masses $m_c$ and VB effective masses $m_v$ (here given as positive values as in the hole picture) and the corresponding relative static dielectric constant $\epsilon_r$ and exciton Bohr radius $a_B^{2D}$ for MoS$_2$ and MoSe$_2$ Monolayer (see text).}
\begin{center}
\begin{tabular}{l c c c c c c c c c c c c c }
\hline 
\hline
&  & $E_g$ & & $E_b$ & & $m_c$ & & $m_v$ & &$\epsilon_r$ & & $a_B^{2D}$ \\
in &  & eV & & eV & & $m_e$ & & $m_e$ & & - & & \AA \\
\hline
MoS$_2$ & & 2.31 & & 0.58 & & 0.40 & & 0.46 & & 4.47 & & 2.77 \\
MoSe$_2$ & & 2.13 & & 0.51 & & 0.49 & & 0.52 & & 5.17 & & 2.40 \\


\hline
\hline 
\end{tabular}
\end{center}
\label{tab1}
\end{table}
\subsection{Exciton States}
\label{sec:exst}
Our target is now to estimate how far the exciton state extends in k-space around the K$^+$ point. This allows us to estimate the possible impact of the proximity of the local CB minimum at $\Lambda$ on the optical transitions and the comparatively weak valley polarization in ML MoSe$_2$. This order of magnitude discussion is graphically represented in the right column of Fig.~\ref{fig:fig3}. Starting from the standard 2D Wannier-Mott model \cite{Klingshirn:2006a}, the exciton ground state energy writes simply as: 
\begin{equation}
E_{b}=\frac{2 \mu e^{*4}}{\hbar^2},
\end{equation}
with $\displaystyle e^{*2}=\frac{e^2}{\epsilon_r}=\frac{q_e^2}{4 \pi \epsilon_0 \epsilon_r}$.
An estimate of the corresponding exciton Bohr radius is thus given by
\begin{equation}
E_{b}a_{B}^{2D}=\frac{e^2}{2\epsilon_r}
\end{equation}
Combining these equations allows for the calculation of the relative dielectric constant, knowing the binding energy $E_b$ and the reduced mass of the electron-hole pair, extracted from $G_0W_0$+BSE
calculations. It yields values of 4.47 and 5.17 for MoS$_2$ and MoSe$_2$, respectively. The main differences from previous theoretical results arise from our choice of a 2D model, the use of our own computed $E_b$ and $\mu$.
These values are usually larger than the estimates found in Ref. \onlinecite{Cheiwchanchamnangij:2012a, Ramasubramaniam:2012a} using  $\displaystyle \kappa=\sqrt{\epsilon_{\perp}\epsilon_{\parallel}}$ (3.44) or by our own direct calculation of the static dielectric constant at the DFT level including local field effect (2.61) for MoS$_2$.  Using the same $\kappa$ approximation we find 2.80 for MoSe$_2$ in comparison with 5.17 of Table \ref{tab1}.\\
\indent Continuing our discussion based on a simple hydrogen-like model, the exciton ground-state wave function can be written as:
\begin{equation}
\phi_{1s}(r)=\sqrt{\frac{2}{\pi a_0^2}}~\textrm{e}^{-r/a_0},
\end{equation}
where $a_0 = 2 a_B^{2D}$ and $r$ is the coordinate of the relative electron-hole motion. The corresponding Fourier transform of this ground state provides a rough estimation of how far around the K$^{+}$ point the exciton state is spread in k-space.
\begin{equation}
\phi_{1s}(q)\approx \frac{1}{\left(1+\left(2~q~a_B^{2D}\right)^2\right)^{3/2}},
\end{equation}
Our representation of $\phi_{1s}(q)$ for ML MoSe$_2$ in Fig.~\ref{fig:fig3} suggest the following scenario: The close proximity of the $\Lambda$ conduction band minimum could be at the origin of the low valley polarization degree of the A-exciton 1s state luminescence, as contributions to the exciton wavefunctions away from the K$^\pm$ points do not obey the strict chiral valley selections rules. It is important to underline in Fig.~\ref{fig:fig3} the striking difference with ML MoS$_2$, where the exciton state is mainly build with K$^+$ electronic states as the contribution from the states around the $\Lambda$ point at much higher energy is negligible. 
\section{Conclusions}
\label{sec:concl}
We probe the exciton states in ML MoSe$_2$ in 1- and 2-photon PLE. We see a clear signature of the 2p state of the A- and B-exciton about 180~meV above the respective 1s exciton state in 2-photon PLE. Our post-DFT calculations reveal in the MoSe$_2$ ML conduction band a local minimum at the $\Lambda$ point only 46~meV above the global minimum at the K-points. As excitons with large binding energies ($E_b\approx0.5$~eV) involve k-states far away from the K-point, the contribution of states around the $\Lambda$ point to optical transitions is possible. As here the chiral optical selections rules are not applicable, these states could contribute to the overall low valley polarization reported for MoSe$_2$ \cite{Li:2014a,Macneill:2015a,Wang:2015a}. To further test this hypothesis, experiments that modify the band structure via strain tuning \cite{He:2013a,Wang:2013c} would provide useful information. Also a systematic study of valley polarization in MoS$_2$Se$_2$ alloy monolayers \cite{Mann:2014a} will help clarifying, why MoS$_2$ shows strong and MoSe$_2$ shows very weak valley polarization in non-resonant PL experiments. 
\section*{Acknowledgements}
\indent We thank Misha Glazov for fruitful discussions and acknowledge funding from ERC Grant No. 306719, ANR MoS2ValleyControl and Programme Investissements d'Avenir ANR-11-IDEX-0002-02, reference ANR-10-LABX-0037-NEXT. I. C. Gerber thanks the CALMIP initiative for the generous allocation of computational times, through the project  p0812, as well as the GENCI-CINES and GENCI-CCRT for the grant x2014096649. I. C. Gerber also acknowledges the CNRS for financial support. 
\section{Appendix}
\label{sec:app}
\subsection{Samples and Experimental Set-up}
\label{sec:exp}
\indent MoSe$_2$ ML flakes are obtained by micro-mechanical cleavage of a bulk MoSe$_2$ crystal on an SiO$_2$/Si substrates using viscoelastic stamping \cite{Gomez:2014a}. The ML region is identified by optical contrast and very clearly in PL spectroscopy. Experiments at T=4 K are carried out in a confocal microscope optimized for polarized PL experiments \cite{Wang:2014b}. The MoSe$_2$ ML is excited by picosecond pulses generated by a tunable frequency-doubled optical parametric oscillator (OPO) synchronously pumped by a mode-locked Ti:Sa laser. The typical pulse and spectral width are 1.6 ps and 3 meV respectively; the repetition rate is 80 MHz. The laser average power is tunable from 2 to 200~$\mu$W. The detection spot diameter is $\approx1\mu$m, i.e. considerably smaller than the ML diameter. For time integrated experiments, the PL emission is dispersed in a spectrometer and detected with a Si-CCD camera. 
\subsection{Computational Details}
\label{sec:set}
\indent The atomic structures, the quasi-particle band structures and optical spectra are obtained from DFT calculations using the VASP package \cite{Kresse:1993a,Kresse:1996b}. PBE functional \cite{Perdew:1996a} is used as approximation of the exchange-correlation term. It uses the plane-augmented wave scheme \cite{blochl:prb:94,kresse:prb:99} to treat core electrons. Fourteen and six electrons for Mo and S, Se respectively are explicitly included in the valence. All atoms are allowed to relax with a force convergence criterion below $0.005$ eV/\AA. After primitive cell relaxation, the optimized lattice parameters are 3.22 and  3.32 \AA~for MoS$_2$ and MoSe$_2$ respectively, these values being in good agreement (~1\%) with previous studies\cite{Ramasubramaniam:2012a,Horzum:2013a} and slightly larger than the bulk experimental values. To sample the Brillouin zone a grid of 12$\times$12$\times$1 k-points has been used, in conjunction with a vacuum height of 17 \AA, to take benefit of error's cancellation in the band gap estimates~\cite{Huser:2013a}. This provides exciton binding energies in reasonable agreement with experiments as suggested in different works~\cite{Klots:2014a, MolinaSanchez:2013}. A gaussian smearing with a width of 0.05 eV is used for partial occupancies, when a tight electronic minimization tolerance of $10^{-8}$ eV is set to determine with a good precision the corresponding derivative of the orbitals with respect to $k$ needed in quasi-particle band structure calculations. Spin-orbit coupling was also included non-self-consistently to determine eigenvalues and wave functions as input for the full-frequency-dependent $GW$ calculations~\cite{Shishkin:2006a} performed at the $G_0W_0$ level. The total number of states included in the $GW$ procedure is set to 600, after a careful check of the direct band gap convergence smaller than 0.1 eV.  
We have used the WANNIER90 code~\cite{Mostofi:2008} and the VASP2WANNIER90 interface~\cite{franchini:2012} to interpolate the band structures on a finer grid. Optical absorption spectra have been calculated using Bethe-Salpeter Equation in the Tamm-Dancoff approximation, including the six highest valence bands and the eight lowest conduction bands \cite{Wang:2015b}.


\end{document}